\documentclass[12pt,referee]{aastex}
\usepackage{emulateapj5}
\newcommand{\mincir}{\raise
-2.truept\hbox{\rlap{\hbox{$\sim$}}\raise5.truept\hbox{$<$}\ }}
\newcommand{\magcir}{\raise
-2.truept\hbox{\rlap{\hbox{$\sim$}}\raise5.truept\hbox{$>$}\ }}
\newcommand{\minmag}{\raise
-2.truept\hbox{\rlap{\hbox{$<$}}\raise6.truept\hbox{$<$}\ }}
\newcommand{\be}{\begin{equation}}
\newcommand{\ee}{\end{equation}}
\newcommand{\ba}{\begin{eqnarray}}
\newcommand{\ea}{\end{esqnarray}}
\newcommand{\brr}{\begin{array}}
\newcommand{\err}{\end{array}}
\newcommand{\bc}{\begin{center}}
\newcommand{\ec}{\end{center}}

\newenvironment{inlinefigure}{%
\def\@captype{inlinefigure}%
\noindent\begin{minipage}{\linewidth}\begin{center}}
{\end{center}\end{minipage}\smallskip}
\makeatother

\shorttitle{XMM - SDSS clusters}
\shortauthors{Plionis et al.}

\begin{document}

\title{XMM-{\em Newton} Observations of Optically Selected SDSS Clusters}

\author{M. Plionis$^{1,2}$, S. Basilakos$^{1}$,
I. Georgantopoulos$^{1}$, A. Georgakakis$^{1}$}
\affil{$^{1}$Institute of Astronomy \& Astrophysics, National Observatory of
Athens, I.Metaxa \& B.Pavlou, P.Penteli 152 36, Athens, Greece \\
$^{2}$ Instituto Nacional de Astrofisica, Optica y Electronica (INAOE)
 Apartado Postal 51 y 216, 72000, Puebla, Pue., Mexico}

\begin{abstract}
We explore the X-ray properties of a subset of the optically selected 
 SDSS cluster sample of Goto et al. (2002), 
 by analysing seven public XMM-{\em Newton} pointings, with exposure times
ranging from $\sim$ 4 to 46 ksec.  
There are in total 17 SDSS
clusters out of which only eight are detected at X-ray wavelengths
with $\rm f_{0.5-2keV}\magcir 1.2 \times 10^{-14}$ ergs cm$^{-2}$ s$^{-1}$.
For the remaining 9 SDSS clusters we estimate
their 3$\sigma$ luminosity upper limits (corresponding to 
$L_x\mincir 5\times 10^{42}$ ergs/sec in the 0.5-2 keV band). 
This relatively low luminosity
suggests that if real structures, these galaxy aggregations
correspond to poor groups of galaxies.
Using the SDSS photometric catalogue we also derive the 
cluster optical $r$-band luminosities. The resulting
scaling relations ($L_{\rm opt}-L_x$, $L_{\rm opt}-T_x$)
are consistent with those of other recent studies.
\end{abstract}


\section{Introduction}
The intense recent interest on galaxy cluster studies is based on the
tight cosmological constraints that they can provide, not only from 
their large-scale distribution, baryon fraction, clustering and
dynamics, but lately also from their internal dynamics, its evolution and 
non-linear physical processes, that give rise 
in a variety of phenomena related to their formation processes 
(eg. Peebles et al. 1989;  Borgani \& Guzzo 2001; Rosati, Borgani \&
Norman 2002). However, in order for such studies to provide the tight 
constraints on cosmological parameters, that we have recently 
seen from the amazing CMB experiments (cf. {\sc Boomerang,
Arceops, Wmap} etc) it is essential to understand and eliminate all 
possible systematic biases that enter in the construction 
and identification procedures of cluster samples (eg. Nichol 2002). 

To this end a large effort has recently been put forward to
construct large, bias-free, samples of clusters of galaxies spanning
a wide redshift range (eg. Postman et al. 1996; Olsen et
al. 1999; Gladders \& Yee 2000; Goto et al. 2002; Bahcall et al. 2003). 
A striking result emerging from such studies is that the
identified clusters strongly depend not only on the detection 
algorithm but also on the wavelength used. Optical selection methods,
based on galaxy overdensities in two-dimensions are biased due to projection
effects, although more sophisticated methods, using color
information and/or structural cluster features, provide
a more uniform selection of clusters (cf. Postman et al 1996). 
However, even such algorithms may suffer from projection
effects and biases towards more evolved galaxy populations. 
Alternatively, X-ray selected cluster samples are 
less prone to biases due to the centrally peaked X-ray emitting 
cluster core which
can be seen even at large ($z\sim 1$) distances 
(eg. Stocke et al. 1991; Castander et al. 1995; Ebeling 
et al. 1996a, 1996b; Scharf et al. 1997; Ebeling et al. 2000; 
B\"{o}hringer et al. 2001; Gioia et al. 2001; 
Rosati, Borgani \& Norman 2002). Nevertheless even such identification
procedure could be biased towards the most virialized (and hence more 
X-ray luminous) clusters.  
A novel cluster detection approach has recently been 
proposed using simultaneous multiwavelength data with the use of
Virtual Observatory data (Schuecker et al. 2004). 
A generic problem, however, arises from our ignorance of the
cluster formation and evolution details which further
complicates the characterization of clusters found at high redshifts
and their use as cosmological probes.

Attempts to understand the systematic biases that enter in 
cluster selection procedures have been recently presented in 
the literature
(cf. Donahue et al 2002; Basilakos et al 2004 and references
therein). This letter further contributes to this issue
by investigating the X-ray properties of a subsample of
the Goto et al (2002) clusters detected in the {\sc Sloan Digital Sky
Survey} (SDSS) with the Cut and
Enhance method (CE), which is based on galaxy colors\footnote{The CE
  catalogue is available from T.Goto, but it can also be found in 
http://www.astro.noa.gr/xray/data/catalogues.html}.
We correlate the positions of these CE cluster candidates with the 
extended X-ray source position, in 
order to investigate which of the clusters show significant 
X-ray emission, and thus have a 
larger probability of being true aggregations of galaxies. 
Throughout this paper we adopt the concordance cosmological model
($\Omega_{\rm m}=1-\Omega_{\Lambda}=0.3$), although all relevant
quantities are partametrized according to $H_{\circ}=100 \;h$ km
s$^{-1}$Mpc$^{-1}$, in order to facilitate comparisons with other
studies.

\section{XMM-Newton Data Analysis}
We use XMM-{\it Newton} archival observations, with a
proprietary period that expired before September 2003, that overlap
with the first data release of the SDSS (DR1; Stoughton et al. 2002). Only
observations that use the EPIC (European Photon Imaging Camera; 
Str\"uder et al. 2001; Turner et al. 2001) cameras as the prime
instrument operated in full frame mode were employed. For fields
observed more than once with the XMM-{\it Newton} we use the deeper of
the multiple observations. A total of 7 fields are used with Galactic
$\rm N_H$ in the range $2-8\times10^{20}\rm \, cm^{-2}$ 
 and PN good time intervals between 4 and 46\,ksec (see Table
 1). Note that two of the fields have clusters as prime targets
(Abell 267 and RX\,J0256.5+0006).
 The X-ray data reduction, source detection and flux estimation are
carried out using methods described in Georgakakis et al. (2004).

In total there are 17 CE optical cluster candidates in 
the region covered by the analyzed XMM-{\em Newton} observations (see Table 2).
We search for extended X-ray sources around the 17 CE clusters, in the 0.5-2\,keV merged
PN+MOS images, where available, using the {\it emldetect} task of {\sc sas}.
We find that extended sources 
are associated with 8 out of the 17 clusters (note that two of these
8 clusters were the prime targets of two of the XMM-{\em Newton} pointings).
 These were visually inspected to exclude the possibility 
 of spurious detections due to their possible association with  
the CCD gaps, hot pixels or the field of view edges. 
Two examples of poor clusters (No 2 and 16 of Table 2) 
are shown in Figure 1, where the 
SDSS $r$-band image is overlayed with the X-ray contours.

Note that 3 out of the 17 CE optical clusters are also 
found in the Bahcall et al. (2003) cluster
candidate list, which is based on the joint application of two
different algorithms on the SDSS photometric galaxy data, while it
contains 2 more clusters not found in Goto et al. Out
of the 3 common clusters, 
two are detected in X-ray's (No.5 and No.13 of our Table 2).

For each cluster candidate we estimate the X-ray flux from a region which 
 encompasses all the extended emission. Then  we correct to the 
 total flux by integrating a King profile to infinity:
$ l_x \propto [ 1+\left(r/r_c\right)^2 ]^{-3 \beta + 1/2}\;\;,$
where $\beta$ is the ratio of energy per unit mass in galaxies to that in gas.
We fix the exponent of the profile to $\beta=0.5$ and the core radius
 to 100 kpc for clusters with (uncorrected) luminosities
$L_x\le 10^{43}\; h^{-2}$ ergs/sec and 200 kpc for those
 with $L_x>10^{43} \; h^{-2}$ ergs/sec. We find that varying the core radius by $\pm50$ kpc 
 changes the total flux by $\sim$ 8 per cent. The conversion from count rates to flux 
 is performed assuming a Raymond-Smith (RS) spectrum with the  
 temperature derived from the spectral fit (see below).          
 For the clusters  that have not been detected we have estimated their
 3$\sigma$ upper limits, by extracting the counts in a 1 arcmin 
 circle around the cluster optical center. Background 
 counts were estimated from nearby regions. Then count rate upper 
 limits are evaluated using the method described in Kraft et al. (1991). 
We convert the count  rates to flux 
 assuming a RS spectrum with a temperature 
 of 2 keV, typical of poor clusters of galaxies. 
 The total flux is estimated using the  King profile above. 
 
We investigate the X-ray spectral properties of the 
 eight detected clusters by performing individual spectral fittings 
 using the {\sl XSPEC} v11.2 package.
 For four clusters which have enough photon statistics 
 (5, 11, 16, 17) we have used the  standard $\chi^2$ analysis. 
 For the other four we use the C-statistic instead (Cash 1979)
 which is proper for fitting spectra with a limited number of counts. 
 We fit simultaneously the PN and MOS data in the energy range 0.3-8 keV.
 We use a RS spectrum  with the abundance fixed at $Z=0.3$,
 absorbed by the Galactic hydrogen column density listed in Table 1.

Finally we present in Table 2 
the optical and X-ray properties  for our clusters. 
Specifically, we list their optical equatorial coordinates (J2000), 
the detected flux (bold) or the $3\sigma$ upper 
flux limits in units of $\rm 10^{-14} erg \; cm^{-2} \;s^{-1}$, 
the cluster redshift (spectroscopic or photometric estimated by the 
CE method), optical $r$-band luminosities (see next section),
X-ray luminosities ($\rm erg~s^{-1}$) in the 0.5-2 keV band, 
best-fit temperatures and $\chi^2$ values, where available, 
of the spectral fits.

It is interesting that the majority of the CE optical cluster
candidates are not identified
in X-rays which could be either because they are poor,
dynamically young, systems which have very low X-ray
emission, or the result of projection effects. 
The 3$\sigma$ upper X-ray luminosity limits are
low, for most cases $\rm L_{0.5-2 keV}<3\times 10^{42}\; h^{-2}$ $\rm erg~s^{-1}$, 
 corresponding to the emission of poor groups of galaxies.
In order to investigate this issue we  determine the optical
properties of all 17 CE cluster candidates.

\section{Optical SDSS data analysis}
We estimate the optical luminosity of the cluster candidates 
by cross-correlating the cluster positions with the SDSS photometric 
$r$-band galaxy survey. For each cluster we select all galaxies with
$m_r\le 21$ (a limit to which there is excellent star-galaxy
separation) falling within a distance of 0.5 $h^{-1}$ Mpc
from the cluster center (for the cases where no spectroscopic redshift
was available we used the estimated Goto et al. photometric
redshift). However, the cluster galaxy membership is
affected by foreground/background contamination and therefore we
estimate the local background by using a circular 
area of a 1.5 degrees radius around each cluster center. 
We first exclude the high surface density regions and then measure
the background surface density, $\sigma_b$. The background galaxy 
counts, projected on the surface covered by each cluster at its 
rest-frame, are estimated by: $N_{\rm back}= \pi d^{2} \sigma_b$, 
where $d$ is the angular cluster radius corresponding to 0.5 $h^{-1}$
Mpc at the cluster rest-frame.
Then the predicted number of the real members
for each particular SDSS cluster is found by:
\be\label{eq:N1}
N_{\rm r}=N_{\rm obs}-N_{\rm bac}\;.
\ee
We assume that for each cluster the luminosity 
function of galaxies can be described by a double Schechter 
function (eg. Smith et al. 1997), ie., 
$\Phi(L)=\phi_{*}[f_{1}(L)+2f_{2}(L)]$, with
\be 
f_{i}(L)=\frac{1}{L_{*,i}}
\left( \frac{L}{L_{i, *}}
\right)^{\alpha_{i}}
{\rm exp} \left( -\frac{L}{L_{*,i}}
\right) \;\;\; i=1,2
\ee
where $L_{*,1}\simeq 3.21\times 10^{11} \;h^{-2} L_{\odot}$ with
$\alpha_{1} \simeq -1.0$ and 
$L_{*,2}\simeq 2.02\times 10^{10} \;h^{-2} L_{\odot}$ with
$\alpha_{2}\simeq -1.7$, respectively. Thus, the number of cluster
galaxy members $N_{\rm r}$ can also be written:
\be\label{eq:N2}
N_{\rm r}= V_{\rm clus} \int_{L_{\rm min}(r)}^{\infty} \Phi(L){\rm d}L \;\;,
\ee
where $L_{\rm min}(r)$ is the minimum luminosity of a
galaxy at distance $r$ (corresponding to the flux limit), $r(z)$
is the luminosity distance and
$V_{\rm clus}$ is the volume occupied by the cluster.

From eq.(\ref{eq:N1}) and eq.(\ref{eq:N2}) we can determine for
each cluster the unknown value of normalization constant $\phi_{*}$ and 
thus obtain the total cluster luminosity by:
$L_{\rm opt}=V_{\rm clus} \int_{L_{\rm min}(r)}^{\infty} L\Phi(L){\rm d}L$.
Combining the above system of 
equations we obtain the final expression for $L_{\rm opt}$:
\be 
L_{\rm opt}= N_{\rm r} 
\frac{\int_{L_{\rm min(r)}}^{\infty} L \; [f_{1}(L)+2f_{2}(L)] {\rm d}L}
{\int_{L_{\rm min(r)}}^{\infty} \left[f_{1}(L)
 +2 f_{2}(L) \right] {\rm d}L}\;\;.
\ee  
The optical cluster luminosities and their 2$\sigma$ Poisson uncertainties
are presented in Table 2. We find that one cluster, also undetected in X-rays, has 
a negative $N_r$  value and is therefore suspect of being a fluke.

Under the assumption that both, cluster gas and galaxies are in hydrostatic equilibrium with the
underline gravitational potential and also 
that the clusters have an approximately constant $M/L$,
one expects power-law scaling relations between $L_{\rm opt}$, $L_x$ 
and $T_x$. 
Although our sample is quite small we 
present as an example in Figure 2 the $L_{\rm opt} -
L_x$ relation and the best fit log-log relation (solid line). The
dashed lines encompass the corresponding Popescu et al. (2004) relation. 
The cluster detections are shown as solid points and those with only upper
limits in $L_x$ are shown as vectors, while we also plot
their 2$\sigma$ $L_{\rm opt}$ uncertainties. Note that the cluster with the
highest 3$\sigma$ upper $L_x$ limit (No 8 in Table 2) is located on
the XMM-{\em Newton} pointing with the lowest exposure time (3.8 ksec)
and thus we are unable to provide tight constraints on its X-ray
properties.

Performing a linear regression in log-log space between the different 
observational quantities we find the following relations:
$L_{\rm opt}/L_{\odot}=10^{11.60(\pm 0.10)} \; L_x^{0.43(\pm 0.13)}$ 
and
$L_{\rm opt}/L_{\odot}=10^{10.80(\pm 0.15)} \; T_x^{1.00(\pm 0.31)}$,
with excellent fits ($\chi_{\nu}^2 \simeq 0.97$ and 0.7 respectively).
The above fits are in good agreement (within the 1$\sigma$ errors)
with the Popescu et al (2004) relations, derived
from a cross-correlation of the RASS and the SDSS (114
clusters). Furthermore, they are in relatively good
agreement with the theoretical expectations for virialized structures
($L_{\rm opt} \propto L_x^{0.5}$ and $L_{\rm opt} \propto
T_x^{1.5}$).

\section{Conclusions}
In order to investigate the X-ray emission of the Goto et al. (2002)
SDSS clusters, we have analysed seven public 
XMM-{\em Newton} pointings with exposure times ranging from 4 to 46 ksecs.
Out of the 17 such SDSS clusters, 
found in the regions covered by these XMM-{\em Newton} pointings,
only eight ($\sim 47\%$) are found to have extended X-ray emission
with limiting flux $f_x \sim 1.2 \times 10^{-14}$ ergs cm$^{-2}$
s$^{-1}$. Note that two of these eight SDSS clusters, which are
also the most X-ray luminous ones, were the prime 
targets of two of the XMM-{\em Newton} observations analysed.
The remaining 9 clusters are probably either very poor and/or
dynamically young clusters with
weak X-ray emission or the results of projection effects. 
The derived $L_{\rm opt}-L_x$ and $L_{\rm opt}-T_x$ relations, 
using the clusters with detected X-ray emission, are consistent 
with those of other recent determinations. 

The present work provides a first glimpse of the 
observations that the future {\sc Duo} X-ray mission will
perform. It will survey a wide angle (6000 sq. deg) in the 
0.3-10 keV energy band with the aim of using $\sim$ 10000 clusters to map 
the geometry of the Universe and determine with high precision its
Dark Energy component. {\sc Duo} 
will target areas covered by the SDSS in order 
to facilitate the cluster redshift determination. 
Our work helps to determine the  X-ray flux depth required 
to identify a significant fraction of the optically selected cluster 
population.

\acknowledgments
This research was jointly funded by the European Union
and the Greek Government in the framework 
of the project {\em 'X-ray Astrophysics with ESA's mission XMM-{\em Newton}'}.
within the program 'Promotion
of Excellence in Technological Development and Research' and
by the Greek-British scientific bilateral program (2002-2003) 
entitled {\em ``Observations of clusters and groups of galaxies with
  XMM-{\em Newton}''}. We thank the XMM-{\em Netwon} science 
archive for the provision of the X-ray data used.


\newpage

\begin{inlinefigure}
\plottwo{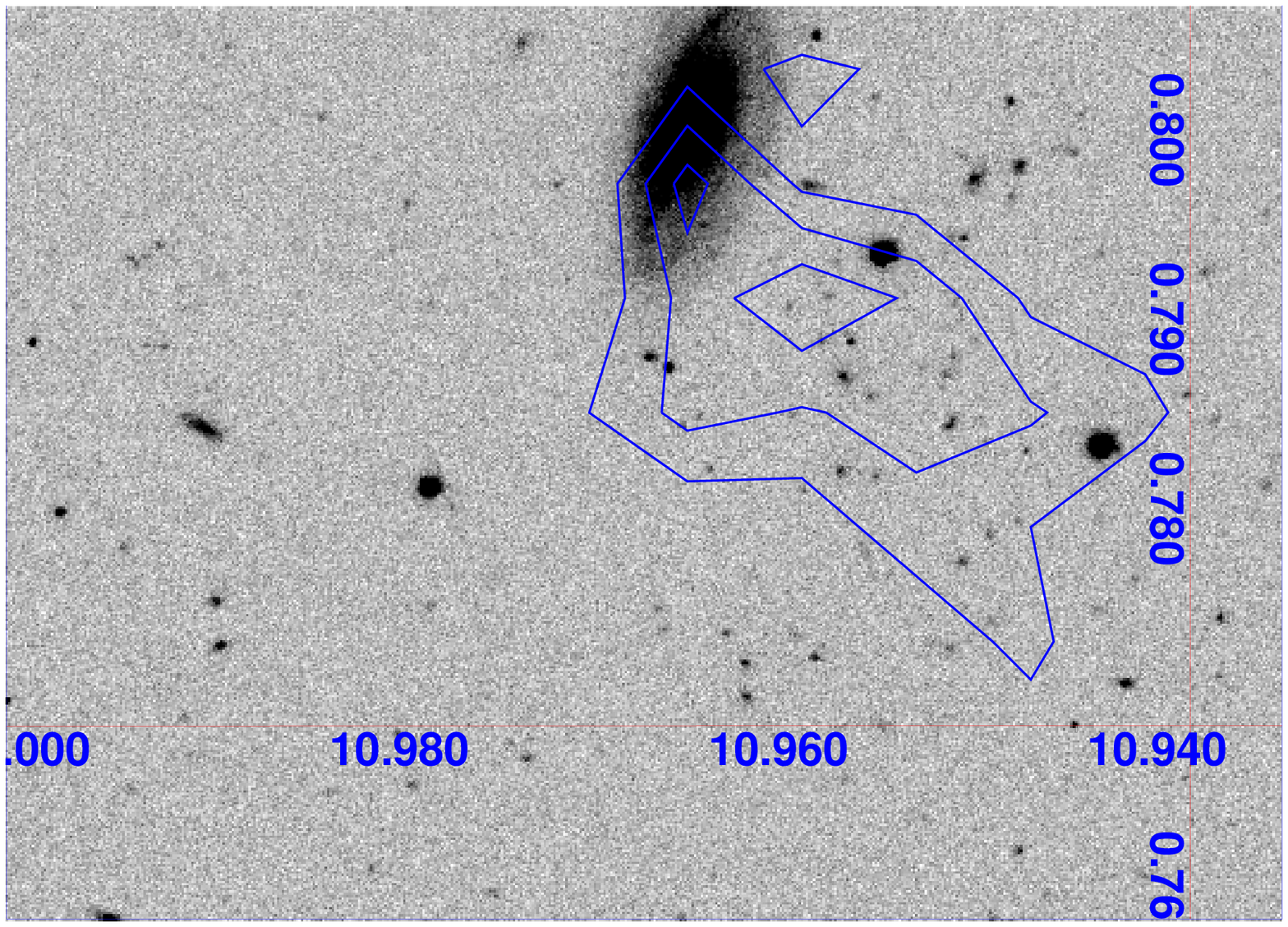}{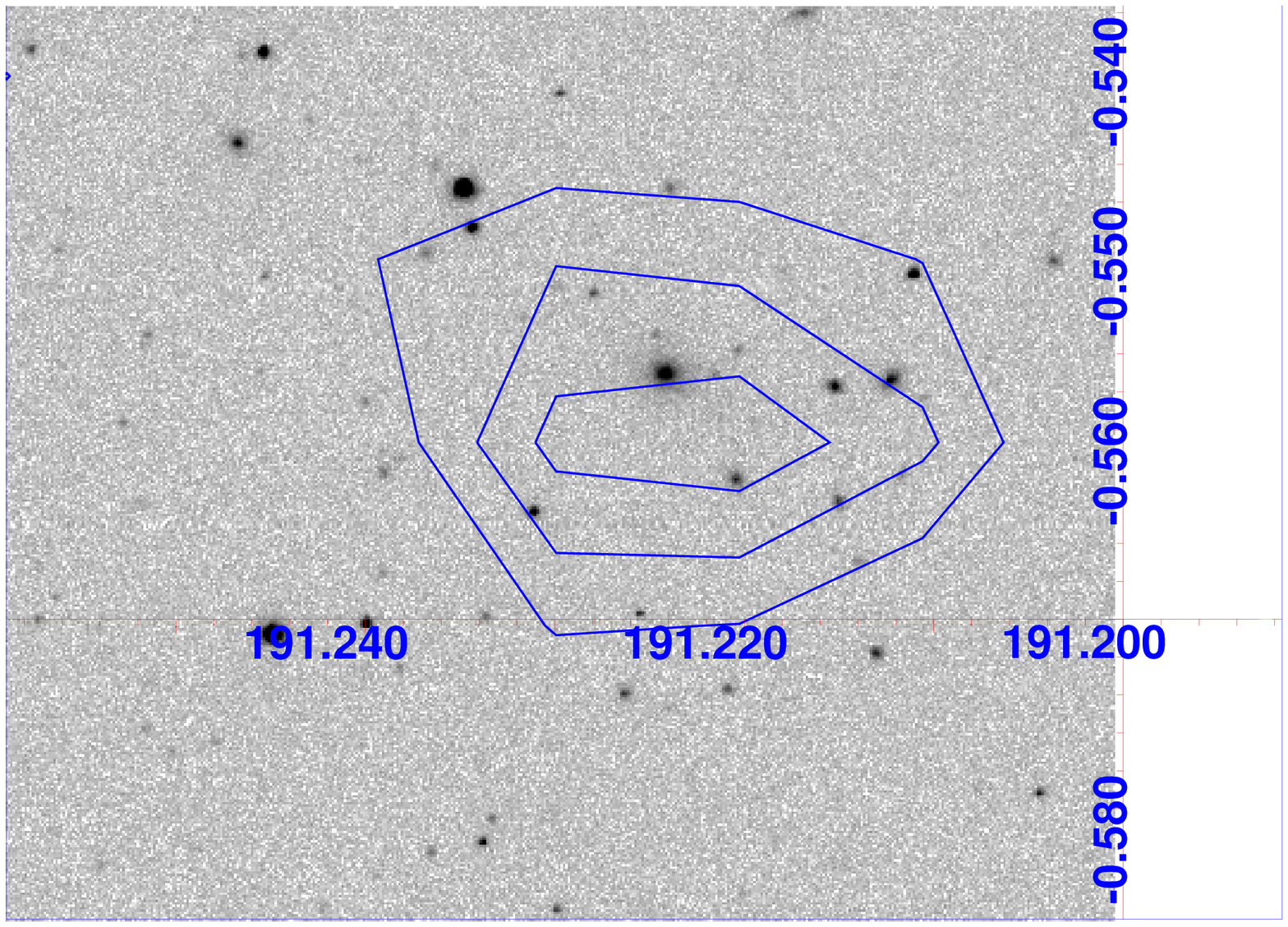}
\figcaption{Two of the Goto et al. (2002) CE clusters detected also in
  X-rays. SDSS $r$-band images are overlayed with X-ray contours for 
the clusters No 2 and 16 of Table 2.}
\end{inlinefigure}

\begin{inlinefigure}
\epsscale{1.0}
\plotone{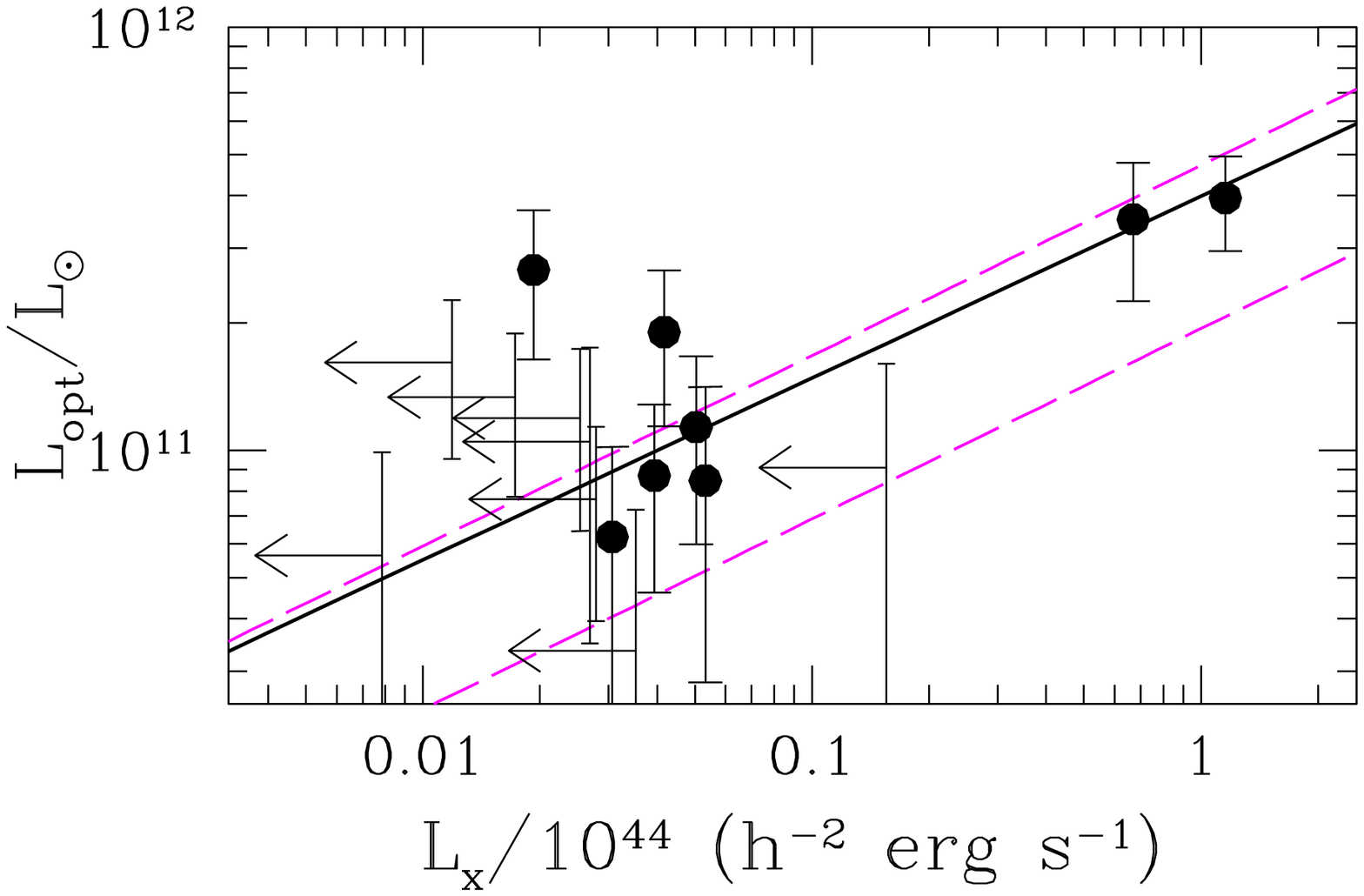}
\figcaption{The $L_{\rm opt}-L_x$ relation from the cluster X-ray
  detections. The dashed lines delineate the Popescu et al. (2004)
  results, while the
thick continuous line corresponds to our best fit. We also show the
clusters with upper $L_x$ limits (vectors).}
\end{inlinefigure}

\begin{table*}
\footnotesize
\tabcolsep 4pt
\caption{The archival XMM-{\it Newton} pointings used in this study.}\label{log}
\begin{center}
\begin{tabular}{ccccccccl} \hline
 RA  & Dec  & Filter & $\rm N_H$  & PN exp. & MOS1 exp. &  Field name & Obs. ID & P.I.\\
(J2000) & (J2000) & & ($\rm 10^{20}\,cm^{-2}$)  & (sec) &  (sec) &  &
 & \\ \hline
00 43 20 & --00 51 15 & Medium & 2.33 & 15\,700 &  --     & UM\,269 &
 0090070201 & Reeves J.\\
01 52 42 &  +01 00 43 & Medium & 2.80 & 5\,800  & 17\,200 & ABELL\,267
 & 0084230401 & Kneib J.P.\\
01 59 50 & --00 23 41 & Medium & 2.65 & 3\,800  &  --     & MRK\,1014
 & 0101640201 & Aschenbach B. \\
02 56 33 & --00 06 12 & Thin   & 6.50 &  --     & 11\,600 &
 RX\,J0256.5+0006 & 0056020301 & Arnaud B.\\
03 02 39 & +00 07 40 & Thin   & 7.16 & 38\,100 & 46\,900 & CFRS\,3H &
 0041170101 & Gear W.	\\
03 38 29 & +00 21 56 & Thin  & 8.15 & 8\,900  & 6\,700  &
 SDSS\,033829.31+00215 & 0036540101 & Brandt W.\\
12 45 09 & --00 27 38 & Medium & 1.73 & 46\,300 & 55\,500 & NGC\,4666
 & 0110980201 & Jansen F.\\ \hline
\end{tabular}
\end{center}

\end{table*}

\begin{table}
\caption[]{X-ray flux upper limits and 
detections (in bold) of the Goto et al. (2002) CE cluster candidates. 
Photometric and spectroscopic redshifts are indicated by the
superscript 1 and 2, respectively. The second column lists the
corresponding number of the Bahcall et al. (2003) SDSS clusters.}
\tabcolsep 6pt
\begin{tabular}{cccccccccc} \\ \hline \hline
$\#$ & $N_{\rm Bahcall}$ &$\alpha$ & $\delta$ & $f_{x}$ & $z$ & $L/L_{\odot}(\times 10^{11})$&${\rm log}L_{x}$ & $\rm T (keV)$ & $\chi_{\nu}^2$ \\ \hline
1& -   &{\bf 10.89}&   {\bf 1.015}& {\bf 8.3}&   {\bf 0.195$^{1}$}& {\bf 0.9 $\pm 0.4$}&{\bf 42.60} & {\bf 1.3$^{+0.3}_{-0.2}$}&  \\
2& -   &{\bf 10.96}&   {\bf 0.774}& {\bf 3.7}& {\bf  0.311$^{2}$}& {\bf 0.9 $\pm 0.6$} &{\bf 42.72} & {\bf 1.4$^{+0.5}_{-0.2}$}&  \\
3& -   &{\bf 10.72}&   {\bf 0.726}& {\bf 4.3}&  {\bf 0.265$^{2}$} &{\bf 1.9 $\pm 0.8$}&{\bf 42.62} & {\bf 2.9$^{+7.6}_{-1.4}$}&  \\ 
4& 173$^a$ &28.05& 1.108& 1.7&  0.243$^{2}$&  $1.6\pm 0.7$&42.07 & & \\
5& 174 & {\bf 28.18}&   {\bf 0.999}& {\bf 174}& {\bf 0.230$^{1}$}& {\bf 4.0$\pm 1.0$}&{\bf 44.06}& {\bf 6.9$^{+0.3}_{-0.2}$} & 997/768  \\ 
6& -   &28.18& 1.138& 2.7&   0.219$^{2}$& $1.3\pm 0.6$&42.24 & & \\
7& -   &28.24& 1.055& 3.6 &0.231$^{2}$ & $1.2\pm 0.6$&42.40&  & \\
8& -   &29.84&   0.449& 2.1 &0.402$^{2}$ & $0.9\pm 0.7$&43.19& & \\
9& -   &29.88& 0.215& 1.5&  0.367$^{2}$& $1.1\pm 0.7$ &42.43 & & \\
10& -  &29.88&   0.283& 1.8&  0.356$^{2}$& $0.3\pm 0.4$&42.55&  & \\
11& -  &{\bf 44.14} & {\bf 0.106}& {\bf 32.4}&  {\bf 0.360$^{1}$}& {\bf 3.5 $\pm 1.3$}&{\bf 43.83} & {\bf 5.6$^{+0.7}_{-0.5}$}& 202/103 \\
12& -  &45.82& -0.005& 0.8&  0.277$^{2}$&$0.6\pm 0.4$ &41.90 & &\\
13& 259& {\bf 54.49}&   {\bf 0.485}& {\bf 1.2}&  {\bf 0.323$^{1}$}& {\bf 2.7 $\pm 1.0$}&{\bf 42.28}& {\bf 1.5$^{+0.7}_{-0.6}$}&  \\
14& - &54.54&   0.274& 6.6&  0.186$^{2}$ &$0.8 \pm 0.4$ &42.45& \\
15& - &191.13&   -0.527& 0.9&  0.231$^{2}$& &41.80 & \\ 
16& - &{\bf 191.22}&  {\bf -0.560}& {\bf 6.9}& {\bf 0.232$^{1}$}& {\bf 1.1$\pm 0.5$}& {\bf 42.70}& {\bf 1.6$^{+0.2}_{-0.2}$}& 128/125  \\
17& - &{\bf 191.22}&  {\bf -0.445}& {\bf 4.3}& {\bf 0.231$^{1}$} &{\bf 0.6$\pm 0.4$} & {\bf 42.50} & {\bf 1.5$^{+0.3}_{-0.2}$}& 168/115  \\
\hline
\end{tabular}

   $^a$ at a distance of $\sim$2.3 arcmin ($\sim 0.5 \; h^{-1}$ Mpc at
   the cluster rest-frame).
\end{table}

\end{document}